\documentclass[sigconf,authorversion,nonacm]{acmart}

\usepackage[unbalanced=true]{idxlayout}
\usepackage{amsmath}

\newcommand{\id}[1]{\mbox{\it #1\/}}
\newcommand{\bid}[1]{\mbox{\bf #1}}

\newcommand{\rid}[1]{\mbox{\rm #1}}
\newcommand{\Exp}{\mbox{\rm E}}
\newcommand{\Var}{\mbox{\rm var}}
\newcommand{\REs}{r.e.\hspace{.19em}}
\newcommand{\RE}{r.e.}

\begin{document}

\title{Quantitative Toolchain Assurance}

\author{Dennis Volpano}
\author{Drew Malzahn}
\author{Andrew Pareles}
\author{Mark Thober}
\affiliation{%
  \institution{Johns Hopkins University Applied Physics Lab}
  \city{Laurel}
  \state{MD}
  \country{USA}
}

\settopmatter{printacmref=false}
\settopmatter{printfolios=true}


\begin{abstract}
The software bill of materials (SBOM) concept aims to include more information
about a software build such as copyrights, dependencies and security references~\cite{ntia}.
But SBOM lacks visibility into the process for building a package.
Efforts such as SLSA \cite{slsa} try to remedy this by focusing on the quality of the build process.
But they lack quantitative assessment of that quality.
They are purely qualitative.
A new form of assurance case and new technique for structuring it, called
{\em process reduction\/}, are presented.
An assurance case for a toolchain is quantitative and when structured as a process reduction can
measure the strength of the toolchain via the strength of the reduction.
An example is given for a simple toolchain.
\end{abstract}

\maketitle

\section{Introduction}

A software toolchain typically contains tools and processes for continuous integration and development.
Tools might include automated testing and/or fuzzing.
Code may be analyzed for undefined behaviors statically, dynamically, or both.
At some point the analyzed executable is packaged and perhaps signed for deployment.
The software bill of materials (SBOM) concept basically extends packaging to include more information
about a build. 
Through standards like Software Package Data Exchange (SPDX), an SBOM affords greater visibility into 
the software's composition through metadata covering libraries used,
copyrights, dependencies and security references~\cite{ntia}.

SBOM lacks visibility into the process used to build a package.
Supply-chain Levels for Software Artifacts (SLSA)~\cite{slsa} tries to remedy this by 
imposing requirements on the process that warrant giving it a security level (0-3).
Level 1 requires provenance be established, Level 2 requires use of a hosted build platform and code
signing, and Level 3 requires
the build platform be hardened with strong tamper protection.
The levels are qualitative.
They are assigned without any quantitative assessment of processes or tools.
For instance, how hard is it for an intruder or impostor to sign untrusted code on behalf of a trusted party?
How strong is tamper protection and just how ``hardened'' is a platform that uses it?
In general, the strength of static analyzers like code sanitizers should be quantified as well.
Merely noting their use to justify assigning a security level is insufficient if they can falsely report security.
More important is knowing how often false reports have occurred.

An assurance case is a structured argument with evidence used to justify a claim that a system will perform
acceptably in a given operating environment~\cite{piovesan2016}.
In industry, these cases are typically organized around the architecture of a system and used
to establish system safety~\cite{nist-ac-2009}.
Lately they have also been used to argue for software pipeline security \cite{chick2023}
but the assurance cases are qualitative just as they are in SLSA.
A new form of assurance case and a new technique
for structuring it, called {\em process reduction\/}, are introduced.
An assurance case for a toolchain is quantitative and when structured as a process reduction can
measure the strength of the toolchain via the strength of the reduction.

\section{Process reductions}

Software toolchains consist of processes following best practices for secure software 
development, as articulated in guidance like NIST SP800-218~\cite{nist800-218}.
For example, there is a process for generating code signing key pairs on a device.
It has its own guidance that states it must be ``sufficiently protected, as the security of this 
process relies upon the protection of the private key'' \cite{nist-code-sign}.
The statement clearly identifies a process whose security is of concern but how does one know
the steps they have taken are enough to ensure security?
What exactly is the definition of security?
To address these questions,
suppose security is the property, say $(A)$, that only the code signer can ever (A)uthenticate as the common name
in the signer's certificate.
Then one can begin to list hypotheses seemingly sufficient to imply the property:
\begin{enumerate}
\item[$(p1)$] Signer is issued an RSA public/private key pair
\item[$(p2)$] Factoring the product of two 1024-bit primes is hard
\item[$(p3)$] Signer's private key is known only to signer
\end{enumerate}
Note $(p2)$ is a critical assumption that must be made explicit if one expects to argue that the property holds.
What makes it unique is that it is beyond the toolchain architect's control.
An architect can introduce steps to increase 
the likelihood that $(p1)$ and $(p3)$ hold.
For instance, for $(p1)$, a RealID may be required to authenticate the signer to the certification authority,
and for $(p3)$, a strong PIN may be required to protect the private key.
But similar steps cannot be taken for $(p2)$.
In fact, it may eventually be disproved.

Though it may seem that $(p1) \wedge (p2) \wedge (p3)\Rightarrow (A)$ is true, it cannot be proved.
Accessing a private key by say a smart card reader might be possible without knowing the PIN in the future
through a new type of reader attack. 
So $(A)$ could be false even when the antecedent holds.
The implication can only be disproved.
But the longer it goes without being disproved, the more widely it is accepted as being true
(a more well-known example of this type of implication is the Church-Turing thesis \cite{HMU2006}).
Suppose then it is accepted as true, though unprovable.
The contrapositive is 
\[\neg(A) \Rightarrow \neg(p1) \vee \neg(p2) \vee \neg(p3)
\]
If $(p1)$ and $(p3)$ are true then $\neg(A) \Rightarrow \neg(p2)$.
The complexity of falsifying $(A)$ rests squarely on the complexity of falsifying $(p2)$.
We say that $\neg(p2)$ {\em process reduces\/} to $\neg (A)$ in this case.
Thus, it is as hard for someone other than the signer to authenticate as the common name as it is to
factor two 1024-bit primes.\footnote{
The same technique can be used to argue for the correctness of a cryptographic protocol where
compromising the protocol should be as computationally hard as compromising its underlying cryptography.}
Note the parallel with many-one reductions.
If problem $L$ many-one reduces to problem $L'$
then $L'$ is as hard as $L$~\cite{kozen2006}.
Unlike many-one reductions, process reductions aren't provable but do have strength.

\subsection{Process reduction strength}

The preceding process reduction is ideal because falsifying $(A)$ rests squarely on falsifying
a hypothesis beyond the architect's control, namely $(p2)$.
It requires proving $(p1)$ and $(p3)$ true in order to eliminate them from the contrapositive.
But neither is provable.
Was it really the trusted signer who was issued the key pair and while the signer may protect the private key
with a PIN, how do we know it wasn't disclosed?
Thus $(p1)$ and $(p3)$ are treated as hypotheses with evidence that may or may not support them.
The strength of the reduction $\neg(p2)$ {\em process reduces\/} to $\neg (A)$ is measured in terms of 
probabilities that $(p1)$ and $(p3)$ are true.
The probabilities can vary because evidence for or against the hypotheses is never final.
So each hypothesis is associated with an independent Beta distributed random variable, a continuous
variable on the interval $[0,1]$.

Beta distributions have three major advantages.
First, a Beta distribution is a conjugate prior for the corresponding Binomial likelihood function in Bayes' rule.
Second, applying Bayes' rule to update a prior based on evidence amounts to merely updating the shape
hyperparameters of the prior, making Bayesian updates fast.
Lastly, for independent Beta distributed random variables $X$ and $Y$, a Boolean logic of Beta distributed random variables
$\neg X$, $X\wedge Y$ and $X\vee Y$ can be defined with multiplicative and additive 
identities $\rid{Beta}(1,0)$ and $\rid{Beta}(0,1)$ respectively \cite{brule2023}.
New random variables can be created to capture logical relationships between hypotheses.
For instance, $p1$, $p2$ and $p3$ are random variables in the code signing example above.
If random variable $p1 \wedge p3$ is associated with $(A)$ then
the strength of the reduction $\neg(p2)\; \id{process reduces to}\; \neg(A)$
is defined to be the mean $\Exp[p1 \wedge p3]$.
Confidence in the strength is defined by the variance $\Var[p1 \wedge p3]$.

The rest of the paper describes techniques for defining random variables for
a simple software toolchain.
An assurance case for the toolchain is given in the form of a process reduction.
The logic and advantages of Beta distributions are detailed in the Appendix.

\section{A simple software toolchain}	\label{simple chain}

Consider a simple toolchain comprising a C compiler, a tool 
(UBSan~\cite{ubsan-tool}) to detect undefined behaviors and a code signing system.
Suppose the compiler is version 8.0.0 of the Clang C compiler, 
and C code is compiled with the option
for detecting undefined behaviors at runtime, specifically option {\tt -fsanitize=undefined}
(we pick version 8.0.0 because it is used in \cite{gemis2019}).
Examples of undefined behaviors include an array subscript exceeding a static bound, dereferencing misaligned or null pointers
and signed integer overflow.
As demonstrated in~\cite{gemis2019}, whether the instrumented code generated with this option detects
an undefined behavior can depend on whether the code has been optimized.
This requires care when defining trials (see Sec.~\ref{trials}).
If Clang and UBSan do not detect any undefined behaviors then all source files are recompiled without the sanitize option
and statically linked without the UBSan runtime libraries.
The executable is then digitally signed using the private key of a key pair whose public key appears in a signer certificate
with common name Alice.

A toolchain comprises stages of steps executed in some order.
Stages are atomic and can be exercised independently of other stages.
The toolchain above has four stages: {\bf compile}, {\bf compile-sanitize}, {\bf compile-code sign}
and {\bf compile-sanitize-code sign}.
Since sanitization and code signing depend on compilation, neither is a stage.
A stage can be thought of as a computational problem with inputs defining instances of the problem.
Every stage has infinitely-many instances, or inputs.
They are C source code files for all stages of the simple toolchain.
Assume {\tt main} bodies have no inputs, as in \cite{gemis2019}, so that
detecting undefined behaviors does not depend on choosing certain inputs.
This restriction can be lifted if the set of inputs remains fixed for each {\tt main} body across toolchain trials.
Otherwise the sanitizer, which is dynamic, can effectively be redefined by varying inputs across trials as different control paths 
afford new opportunities to detect undefined behaviors.
This should be avoided as the aim is to learn about one sanitizer at a time by applying it to different {\tt main} bodies.
See Sec.~\ref{implementation} for details on how to reach a fixed set of inputs, against which sanitizer strength
can be measured, as a fixed point of a function. 

\subsection{Toolchain trials}	\label{trials}

For the purpose of quantifying toolchain assurance,
we must define a trial or experiment.
A {\em trial\/} of a stage in the toolchain is an {\em instance\/} of the stage.
Note that running a stage twice on the same input doesn't constitute two trials.
Just as inputs must differ to be different instances of the same problem, they must differ to be
different trials of the same stage.
In practice, the {\bf compile-sanitize} stage would be repeated until a version of the source file is reached
for which no undefined behaviors are found.
This final version would then be recompiled without the UBSan option but with all other options the same.
The sanitizer step is omitted, however recompiling here is not a trial of {\bf compile}, as the source
code did not change, only the sanitizer option was removed.
In fact, sanitizing the final version is not even a trial of {\bf compile-sanitize} if the signer is
ready to sign the executable obtained by compiling without the sanitize option.
Instead, this compilation would associate with signing and be a trial of {\bf compile-sanitize-code sign}.
Likewise, if linking fails because of an unresolved external reference in a source file
then the compilation fails and there is no new trial of {\bf compile}.
There is also no new trial of it if linking fails due to a problem that does not require 
changing any source code files
(e.g.\ recovering a missing object file by just recompiling an old source file).

Keep in mind that since detecting an undefined behavior can depend on the optimization option chosen
when code is compiled (see page 22 of \cite{gemis2019}), there would be a set of trials for each stage of the toolchain
for the unoptimizing compiler (option -O0) and another set for each stage for the optimizing one (options -O1/2/3).

\section{Assurance case for the toolchain}

An assurance case in the form of a process reduction is a logical implication, widely believed to be true but unprovable,
represented as a tree with the consequent at the root and each node having an 
independent Beta distributed random variable.
Associated with each random variable is a hypothesis.
For instance, an assurance case for the simple toolchain is given in Fig.~\ref{assurance-case}.
It has 8 random variables, each enclosed within
parentheses, and a hypothesis for each.
\begin{figure*}
\begin{center}
\parbox{0cm}{\begin{tabbing}
$(p1)$ \= $(p1)$ \= \kill
$(s1\wedge s2\wedge s3\wedge p1\wedge p3)$ If $Q$ is signed with private key $(d,n)$ then $Q$ 
has no undefined behaviors \hspace{0.5em}$(G)$\\
\> ($s1$) If Alice signs $Q$ then clang -O0/UBSan rejects source used to build $Q$ \\
\> $(s2$) If clang -O0/UBSan rejects source used to build $Q$ then $Q$ has no undefined behaviors \\
\> $(s3)$ Signer of $Q$ is authenticated using $((e,n),(d,n))$ and $Q$ is signed with $(d,n)$ \\ 
\> $(p1\wedge p3)$ Only Alice can authenticate as Alice using $((e,n),(d,n))$ \\
\> \> ($p1$) Alice is issued RSA public-private key pair $((e,n),(d,n))$ with 2048-bit modulus $n$ \\
\> \> ($p2$) Factoring the product of two 1024-bit primes is hard \\
\> \> ($p3$) Private key $(d,n)$ is known only to Alice given $((e,n),(d,n))$ is issued to Alice
\end{tabbing}}
\end{center}
\caption{$\neg(p2)$ process reduces to $\neg (G)$ with strength $\Exp[s1\wedge s2\wedge s3\wedge p1\wedge p3]$}
\label{assurance-case}
\end{figure*}
The hypothesis at the root, namely $(G)$, asserts that if executable $Q$ is signed with private key $(d,n)$
then it is free of undefined behaviors.\footnote{Free variables $Q$, $e$, $d$ and $n$ are implicitly
universally-quantified variables in the tree.}
Hypotheses $(s1)$--$(s3)$ and $(p1)$--$(p3)$ would seem to imply $(G)$.
Only Alice can authenticate as Alice using the RSA key pair by $(p1)\wedge(p2)\wedge(p3)$.
Thus by $(s3)$, Alice signs $Q$ with her private key $(d,n)$. 
If Alice signs $Q$ then by $(s1)$, clang~-O0/UBSan rejects the source used
to build $Q$, which means no undefined behavior was detected.
So $Q$ has no undefined behaviors by $(s2)$.
Therefore we have $\neg(p2)$ process reduces to $\neg(G)$.
In other words, falsifying $(G)$ implies falsifying $(p2)$.
The strength of this reduction is the mean of
the Beta random variable at $(G)$:
\[\Exp[s1\wedge s2\wedge s3\wedge p1\wedge p3]
\]
It requires Beta probability density functions (PDFs) for the constituent random variables.
They in turn rely on evidence of the truth of their hypotheses based on trials.
Thus, they can vary over trials and be updated by Bayesian inference.
In practice, we don't expect the Beta distributions for all variables to vary.
For instance, it would likely be fixed (true or false) over all trials for $p1$.
But we expect it to vary when learning about the completeness of
Clang and UBSan from trials, which is reflected in the Beta PDF for $s2$.

\subsection{Beta PDFs for the random variables}

Suppose we have $m > 0$ trials total of all stages in our toolchain
of which $n > 0$ are trials of stages involving code signing.
Assume each of the $n$ trials requires reauthenticating the Signer as Alice.
Beta PDFs for each of the random variables are defined below.

\begin{itemize}
\item[$s1$] Let $\alpha_{s1}$ be the number of $n$ trials in which Alice signed $Q$
and Clang/UBSan reject the source used to build $Q$, and $\beta_{s1}$ be the number of $n$ trials in which
Alice signed $Q$ and Clang/UBSan accept the source code used to build $Q$.

\item[$s2$] The Beta PDF for this random variable is $\rid{Beta}(\alpha_{s2},\beta_{s2})$.
It measures the completeness of Clang/UBSan over $m$ trials.
It is defined incrementally in practice through Bayesian inference.
The posterior PDF at trial $k-1$ is the Beta PDF used in the 
assurance case for the toolchain at trial $k$, which may undergo
a Bayesian update at trial $k$ for use at trial $k+1$ and so on.
See Sec.~\ref{SA}.
Though updates are optional, performing them more accurately measures completeness.
Initially, $s2\sim \rid{Beta}(19,43)$ (see pg.\ 56 of \cite{gemis2019}).

\item[$s3$] Let $\alpha_{s3}$ be the number of $n$ trials where Alice is authenticated
using the key pair and $Q$ is signed with its private key,
and $\beta_{s3}$ the number of $n$ trials where an impostor of Alice is authenticated
using the key pair and $Q$ is signed with its private key.

\item[$p1$] 
Let $\alpha_{p1}$ be the number of $n$ trials where Alice has been issued the key pair and
$\beta_{p1} = n - \alpha_{p1}$.

\item[$p3$] Let $\alpha_{p3}$ be the number of $n$ trials where Alice is issued the 
key pair and the private key is known only to her,
and $\beta_{p3}$ be the number of $n$ trials where she is issued the key pair but the private key is known to others.
\end{itemize}

\subsection{Reduction strength}

The extent to which falsifying $(G)$ rests squarely on
falsifying $(p2)$ is the strength of the process reduction from $\neg(p2)$ to $\neg(G)$.
The reduction is from $\neg(p2)$ because $(p2)$ is beyond control via any processes in the toolchain.
One could argue that $(p3)$ is also beyond control if Alice does not care about
disclosing the PIN that protects her private key.
In this case, it makes sense to process reduce $\neg(p2) \vee \neg(p3)$ to $\neg(G)$.
The random variable at $(G)$ would now be $(s1\wedge s2\wedge s3\wedge p1)$.
Now the complexity of falsifying $(G)$ rests on that of falsifying $(p2)$ or $(p3)$
with strength $\Exp[s1\wedge s2\wedge s3\wedge p1]$.
Though the strength is greater, it requires admitting in the assurance case that 
Alice is an uncontrolled threat to the PIN.

\section{Beta PDF$\bid{s}$ for program analyses}

A code sanitizer is an example of dynamic program analysis aimed at deciding whether a given program has an undefined behavior,
which is one that is not defined by the semantics of the programming language.
Examples of such behaviors are pointers escaping their scope and array bounds violations.
There are also static program analyses aimed at detecting undefined behaviors.
Some are embedded in compilers like the GCC and Clang C compilers, while others like
Frama-C stand alone \cite{framaC}.
The technique for building a Beta PDF for program analysis is the same whether it is static or dynamic.
Moreover, the technique is independent of the analysis done, as it measures only how complete the analysis is in practice.

The Beta PDF for an analyzer that analyzes programs for a decidable $P$ is easily defined because one can
assume there is a procedure $M$ that accepts $P$ and always halts. 
So for a given program $p$, the Beta distribution becomes $\rid{Beta}(1,0)$ (true) if $M$ accepts $p$
and $\rid{Beta}(0,1)$ (false) if $M$ rejects $p$.
Unfortunately most of the properties of interest to program analyzers are undecidable, or worse, not
even recursively enumerable (\RE).
When the property $P$ of interest to an analyzer is undecidable,
Bayesian inference can be done over trials involving different inputs to measure the analyzer's performance.
For any input, it is possible to see if the analyzer succeeded or failed in determining whether
the input has property $P$ by manually inspecting the input.
Manually inspecting every input is impractical
but suppose it is done to some extent
to learn a Beta distribution for the analyzer.
Then the distribution would be used to extrapolate its performance on inputs not manually inspected.

This approach can be applied to Clang and UBSan.
Suppose $P$ is the set of all C programs with undefined behaviors.
Then $P$ is undecidable.
In proof, a C program can simulate a semi-decision procedure for an undecidable yet \REs set $L$.
Given an input, the program simulates the semi-decision procedure on that input and
exhibits an undefined behavior if the procedure accepts the input.
As the simulation does not require any undefined behaviors, the undefined behavior will be exhibited
only if the procedure accepts the input.
So any algorithm to decide whether a C program has an undefined behavior 
is an algorithm for $L$, which doesn't exist. 

Set $P$ is \REs and Clang/UBSan work together as a procedure to accept it.
Assume they are sound, meaning that if they find an undefined behavior, then one actually exists.
Further, they always halt in practice, so therefore they must be incomplete; otherwise $P$ would be decidable.
Completeness demands that whenever they are given code with an undefined behavior, they detect it.
But this doesn't match what is demanded of them in the assurance case at $(s2)$.
There it states that if Clang and UBSan reject the given source code (find {\em no\/} undefined behaviors)
then the code has none. 
This is a statement about their soundness in accepting the complement of $P$. 
Accepting the complement of $P$ though is not their goal.
However, the soundness statement is the contrapositive of a statement about their completeness 
in accepting $P$, which is their goal.
Therefore what is required of the tools at $(s2)$ can be quantified by 
measuring their completeness in accepting $P$.
This is done using Bayesian inference, as described in the next section.

\subsection{Bayesian updating of $\rid{Beta}(\alpha_{s2},\beta_{s2})$}	\label{SA}

Let $p$ be the source code used to build $Q$ at a trial.
Then the rules for Bayesian updating of $\rid{Beta}(\alpha_{s2},\beta_{s2})$ at the trial are given below:
\begin{enumerate}
\item $\rid{Beta}(\alpha_{s2} + 1, \beta_{s2})$ if Clang/UBSan accept $p$.
\item $\rid{Beta}(\alpha_{s2}, \beta_{s2} + 1)$ if Clang/UBSan reject $p$ and $p$
has an undefined behavior.
\item No update if Clang/UBSan reject $p$ and $p$ has no undefined behaviors.
\end{enumerate}
Rule (1) applies when $p$ has an undefined behavior and Clang/UBSan detect it, which is success.
We know $p$ has an undefined behavior because Clang/UBSan accept $p$ and we assumed they are sound.
Rule (2) applies when Clang/UBSan fail to detect an undefined behavior in $p$.
Rule (3) avoids updating altogether because its facts are actually not informative.
To see this, suppose $p$ has no undefined behaviors.
Then Clang/UBSan cannot accept $p$ because they are sound.
Since they always halt, they reject $p$ and will never do so erroneously because
if they did, it would mean they rejected $p$ when $p$ {\em has\/} an undefined behavior, which it doesn't.
Therefore, Clang/UBSan always succeed when they reject an input that has no undefined behaviors.
In fact, they would always succeed on such inputs even if they did nothing but reject {\em all\/} inputs!
These successes therefore should not count toward completeness even though the facts in rule (3) look like
hypothesis $(s2)$.

\subsection{Test-driven Bayesian updating}	\label{implementation}

Recall from Sec.~\ref{simple chain} the restriction that {\tt main} bodies have no inputs.
As mentioned, the restriction can be lifted if the set of inputs remains fixed for each {\tt main} body across toolchain trials.
This is so for any unit testable component such as a C source file having unique function definitions.
The set of inputs can be defined by test cases for unit and integration testing.
Completness of Clang/UBSan is measured relative to these test cases.
It would be inaccurate to talk about its completeness without them.
First consider unit testing.

Suppose {\em Units\/} is a set of unit testable components and {\em Tests\/} is a set of test inputs for the components.
Not all test cases will be meaningful for every component.
Let $\rid{Beta}(\alpha,\beta)$ be a prior Beta distribution for Clang/UBSan.
Function $H$, defined in Fig.~\ref{unit test}, implements the 3 rules for Bayesian updates given in the previous section
in the context of unit testing.
\begin{figure}
\begin{center}
\[
\begin{array}{l}
H(\id{Units},\id{Tests},\rid{Beta}(\alpha,\beta)) = \\
\hspace{0.5em}((\id{Units}-\id{unit})\cup\{\id{unit}'\}, \id{Tests}, \rid{Beta}(\alpha + 1,\beta)) \\
\hspace{1em}\rid{if}\;\id{unit}\in\id{Units}\rid{,}\;\id{UBSan}\;\rid{detects an undefined behavior when} \\
\hspace{1em}\id{unit}\;\rid{runs on an input in}\;\id{Tests}\rid{,}\;\rid{and}\;\id{unit}'\;\rid{does not have that} \\
\hspace{1em}\rid{instance of the behavior}\\[0.25em]

H(\id{Units},\id{Tests},\rid{Beta}(\alpha,\beta)) = \\
\hspace{0.5em}((\id{Units}-\id{unit})\cup\{\id{unit}'\}, \id{Tests}\cup\{\id{test}\}, \rid{Beta}(\alpha + 1,\beta)) \\
\hspace{1em}\rid{if}\;\id{unit}\in\id{Units}\rid{,}\;\id{UBSan}\;\rid{rejects}\;\id{unit}\;\rid{when it is run on every} \\
\hspace{1em}\rid{meaningful input in}\;\id{Tests}\rid{,}\;\id{unit}\;\rid{has an undefined behavior,} \\ 
\hspace{1em}\id{test}\;\rid{is an input that causes}\;\id{UBSan}\;\rid{to detect the behavior in} \\
\hspace{1em}\id{unit}\rid{, and}\;\id{unit}'\;\rid{does not have that instance of the behavior} \\[0.25em]

H(\id{Units},\id{Tests},\rid{Beta}(\alpha,\beta)) = \\
\hspace{0.5em}((\id{Units}-\id{unit})\cup\{\id{unit}'\}, \id{Tests}, \rid{Beta}(\alpha,\beta + 1)) \\
\hspace{1em}\rid{if}\;\id{unit}\in\id{Units}\rid{,}\;\id{unit}\;\rid{has an undefined behavior, there is no} \\
\hspace{1em}\rid{input to}\;\id{unit}\;\rid{that causes}\;\id{UBSan}\;\rid{to detect the behavior, and} \\
\hspace{1em}\id{unit}'\;\rid{does not have that instance of the behavior} \\[0.25em]

H(\id{Units},\id{Tests},\rid{Beta}(\alpha,\beta)) = (\id{Units},\id{Tests},\rid{Beta}(\alpha,\beta))\;\rid{otherwise}
\end{array}
\]
\end{center}
\caption{Bayesian updates in the context of unit testing}
\label{unit test}
\end{figure}
The first case in the definition of $H$ applies when Clang/UBSan detects an undefined behavior in
{\em unit\/} when run on some input in {\em Tests\/}.
In practice we expect {\em unit\/} will be fixed to remove that instance of the undefined behavior and a new version $\id{unit}'$
will replace it.
The second case is similar except the test input enabling {\em UBSan\/} to detect an undefined behavior in {\em unit\/}
does not belong to {\em Tests\/} and must be created.
The case reflects success of UBSan as long as the new input is added to {\em Tests\/}.
The third case applies when there is no input to {\em unit\/} that enables {\em UBSan\/} to detect an 
undefined behavior in it yet one exists.
Note this covers the case when {\em unit\/} expects no inputs as then there are no inputs to even consider. 
The first two cases implement the first rule in the previous section and
the third case the second rule.

Bayesian updates in the context of integration testing is handled by extending the definition of $H$
as shown in Fig.~\ref{integration test}.
\begin{figure}
\begin{center}
\[
\begin{array}{l}
H(\id{Units},\id{Tests},\rid{Beta}(\alpha,\beta)) = \\
\hspace{0.5em}((\id{Units}-R)\cup R', \id{Tests}, \rid{Beta}(\alpha + 1,\beta)) \\
\hspace{1em}\rid{if}\;R\subseteq\id{Units}\rid{,}\;\id{UBSan}\;\rid{detects an undefined behavior when} \\
\hspace{1em}R\;\rid{runs on an input in}\;\id{Tests}\rid{,}\;\rid{the behavior cannot be detected} \\
\hspace{1em}\rid{through unit testing alone, and}\;R'\;\rid{does not have that} \\
\hspace{1em}\rid{instance of the behavior} \\[0.25em]

H(\id{Units},\id{Tests},\rid{Beta}(\alpha,\beta)) = \\
\hspace{0.5em}((\id{Units}-R)\cup R', \id{Tests}\cup\{\id{test}\}, \rid{Beta}(\alpha + 1,\beta)) \\
\hspace{1em}\rid{if}\;R\subseteq\id{Units}\rid{,}\;\id{UBSan}\;\rid{rejects}\;R\;\rid{when it is run on every} \\
\hspace{1em}\rid{meaningful input in}\;\id{Tests}\rid{,}\;R\;\rid{has an undefined behavior that} \\ 
\hspace{1em}\rid{cannot be detected through unit testing alone,}\;\id{test}\;\rid{is an} \\
\hspace{1em}\rid{input that causes}\;\id{UBSan}\;\rid{to detect the behavior in}\; R\rid{, and} \\
\hspace{1em}R'\;\rid{does not have that instance of the behavior} \\[0.25em]

H(\id{Units},\id{Tests},\rid{Beta}(\alpha,\beta)) = \\
\hspace{0.5em}((\id{Units}-R)\cup R', \id{Tests}, \rid{Beta}(\alpha,\beta + 1)) \\
\hspace{1em}\rid{if}\;R\subseteq\id{Units}\rid{,}\;R\;\rid{has an undefined behavior, there is no} \\
\hspace{1em}\rid{input to}\;R\;\rid{that causes}\;\id{UBSan}\;\rid{to detect the behavior, and} \\
\hspace{1em}R'\;\rid{does not have that instance of the behavior} \\[0.25em]
\end{array}
\]
\end{center}
\caption{$H$ extended for integration testing}
\label{integration test}
\end{figure}
Since trial samples cannot overlap, care is needed when measuring the success of UBSan during integration 
or system testing. 
UBSan should not be credited with success if it detects an undefined behavior during such testing 
that can be detected with unit testing alone.

There is a tuple $(\id{Units},\id{Tests},\rid{BetaPDF})$ prior to testing some version of a system
where $\id{Units}$ is the system under test, $\id{Tests}$ is an initial set of test cases,
and BetaPDF is a prior Beta PDF for Clang/UBSan.
The test cycle for this version is complete with respect to undefined behaviors when a fixed point
of $H$ is computed iteratively.
Let $x_0 = (\id{Units},\id{Tests},\rid{BetaPDF})$, which is in the domain of $H$.
Then the fixed-point iteration of $H$ is
$x_{n+1} = H\;x_n$ for $n=0,1,2,\ldots$,
which is the sequence $x_0,H\,x_0,H\,(H\,x_0),\ldots$ of iterated function applications.
It converges to a fixed point of $H$ if the number of undefined behavior instances 
among systems in the sequence strictly decreases.
UBSan finds no undefined behaviors in the system under test of a fixed point because there are none.
After some version of the system, we expect no more fixed points will be computed because each
requires proving the absence of any undefined behaviors, which may require too much manual effort.
Learning about UBSan is frozen at this point.
Thereafter whenever UBSan finds no undefined behaviors in a new version of the system, relative to tests that
contain at least those of the last fixed point computed, one will take 
the mean of the posterior Beta PDF of this last fixed point as the probability
the system has no undefined behaviors.

While we expect the cost of computing fixed points will diminish as the system matures, it should
be noted that not every test cycle must be completed with a fixed point. 
The tradeoff, however, will be
knowing less about UBSan's completeness.

Note that $H$ will credit UBSan with two successes if it detects the same undefined behavior 
in two different units, or two instances of it in the same unit regardless of the test cases used.
If this behavior were the only type detected by UBSan 
then it might appear $H$ can be ``gamed'' into boosting UBSan's performance,
even when UBSan has limited capability,
if $H$ is always applied to units with only this type of undefined behavior.
But the units to which it is applied come from a real system, call it $S$, and if $S$ comprises units
with this behavior as the only possible type of undefined behavior then $H$ computes a Beta PDF that
accurately reflects UBSan's performance in the context of $S$.
We are not suggesting this performance be used to judge UBSan when building any system using 
the toolchain but rather just when building $S$.
So one could see multiple Beta PDFs for UBSan, one for each project using the toolchain, to reflect
possible differences in its performance across projects.

Finally, as $H$ grows the test suite for a system, UBSan's score improves even though UBSan's
instrumentation algorithm has not changed.
That's fine because new test cases lead to new execution paths to UBSan's instrumentation.
But one could imagine the instrumentation algorithm changing over the lifetime of a toolchain.
The third cases defining $H$ in Figs.~\ref{unit test} and \ref{integration test} handle the case
when no input causes UBSan to detect an undefined behavior.
$H$ could record success here instead of failure if the instrumentation algorithm were adapted
so that the behavior is detected for some input.
If that input were not among the existing test cases then it would be added.
Changes to the algorithm should be monotonic in that all previous successes of UBSan are preserved.

\section{Future work}

A process reduction relies on a community of interest agreeing on a set of conditions {\em sufficient\/} for implying
a desired outcome.
Outcomes are rarely specified with precision in guidelines for secure software development, if at all.
They must be distilled.
The implication will undergo iterations and should eventually converge to something that is
widely agreed upon to hold but can only be disproved.
It becomes a standard at this point and its contrapositive a basis for a process reduction.
A working group could distill sufficient conditions for desired outcomes implicit in
documents like that for code signing \cite{nist-code-sign} and NIST SP800-218 \cite{nist800-218}.
There is precedent for this type of standardization.
A similar effort was undertaken by the Overarching Properties Working Group to identify a
sufficient set of properties for approving airborne systems \cite{holloway2019}.

Software test suites are evaluated in different ways using metrics such as input space coverage, 
code coverage or kill ratios \cite{hamlet2010}.
Assuming these metrics are useful, they should be represented as Beta distributions in our framework.
For instance with mutation testing \cite{jia2011}, one might represent the successes and failures of a test suite by a BetaPDF
where success is determined by whether the suite causes a mutation of
the program under test to exhibit some observable difference in behavior when executed, for instance, 
terminate abnormally \cite{shahriar2008}.
The suite is said to kill the mutant if the difference in behavior is observed.
The shape parameters of a BetaPDF for the test suite in this case would be mutants killed $(\alpha)$
and not killed $(\beta)$ by the suite.

The effectiveness of a test suite, not just a metric for it, 
should also be represented within a toolchain since
the metric may not correlate well with finding faults \cite{holmes2014}.
Correlation should have its own BetaPDF. 
This implies performing Bayesian updates over time to learn how well the metric correlates to finding faults
in a given system, much the way we learned about the completeness 
of Clang/UBSan for a given system over time.
Learning for a class of systems rather than a single system may also be useful.
For instance, Modified Condition/Decision Coverage (MC/DC) \cite{mcdc2001} has long been used as coverage criteria 
for tests involving safety-critical systems.
Experience therefore must have shown that these coverage criteria correlate well with finding faults in such systems.
Either way it may make sense to learn about correlation for specific systems rather than apply the results of
more general empirical studies \cite{holmes2014,chekam2017}.
This is another direction for future work.

\section{Conclusion}

Attacks on software supply chains have heightened awareness of the need to produce evidence
that systems built by them are reliable and safe to execute.
The SBOM described in \cite{ntia}
and qualitative efforts like SLSA \cite{slsa} describe informal
evidence that sits at the least rigorous end of an evidence spectrum.
At the opposite end is the most rigorous evidence,
best characterized by proof-carrying code (PCC) \cite{necula1998}.
Mobile code security was studied extensively in the mid 1990's.
It arose in large part from Java applets and to a lesser extent from DARPA's
Active Networks program where routers could execute small programs within packets.
In each of these cases, one has to guard against malicious executable code.
The idea behind PCC was to couple a proof of some property about an application's executable with the executable.
If the property were consistent with a recipient's security policy then the
recipient would check the validity of the proof, and if valid would execute the code.
Leveraging toolchain facts, we introduce a kind of quantitative evidence
for software that sits somewhere between these two endpoints.

\section{Acknowledgements}

Thanks to Lucja Kot, Greg Nelson and Bill Bierman for references on mutation testing, and
to Elishiva Zak for an update on an effort to evaluate test suites vis-\`{a}-vis mutation testing.

\section{Appendix}

\subsection{Boolean logic of Beta PDFs}	\label{logic}

Let $X$ and $Y$ be two independent Beta distributed random variables.
Each has a mean and variance denoted $\Exp[X],\Var[X]$ and $\Exp[Y],\Var[Y]$.
Random variable $X\wedge Y$ is represented by product $XY$.
As $X$ and $Y$ are independent, $\Exp[XY]=\Exp[X]\Exp[Y]$ and 
$\Var[XY] = \Var[X]\Var[Y] + \Var[Y]\Exp[X]^2 + \Var[X]\Exp[Y]^2$.
Random variable $\neg X$ is represented by variable $1-X$.
Further, $\Exp[1-X] = 1 - \Exp[X]$ and $\Var[1-X] = \Var[X]$.
Lastly random variable $X\vee Y$ is shorthand for $\neg(\neg X\wedge\neg Y)$.
Note mean and variance are undefined for $\rid{Beta}(0,0)$.

A given mean $\mu\in[0,1]$ and nonzero variance $\sigma^2$, uniquely define the PDF hyperparameters
$\alpha$ and $\beta$ of a Beta distribution.
With $\mu = \frac{\alpha}{\alpha + \beta}$ and $\sigma^2 = \frac{\alpha\beta}{(\alpha + \beta)^2(\alpha + \beta + 1)}$,
we have
\[
\begin{array}{ll}
\alpha = \frac{\mu^2 (1-\mu)}{\sigma^2} - \mu & \hspace{1em}
\beta = \frac{\mu(1-\mu)^2}{\sigma^2} - (1-\mu)
\end{array}
\]
The case when $\sigma^2 = 0$ is handled separately.
It arises when representing the identities $\rid{Beta}(1,0)$ and $\rid{Beta}(0,1)$.
If $\sigma^2 = 0$ and $\mu = 1$ then $\alpha = 1$ and $\beta = 0$, otherwise
if $\mu = 0$ then $\alpha = 0$ and $\beta = 1$.
Thus, logical expressions can be evaluated efficiently in terms of mean and variance
and converted to hyperparameters as needed to perform Bayesian updates.
Contrast with \cite{brule2023} where logical terms are evaluated by evaluating 
complex expressions of Beta hyperparameters.

\subsection{Efficient Bayesian updates}	\label{bayes-updates}

Bayesian updates can be done by simply adding to Beta PDF hyperparameters.
To see this, given variables $\pi$ and $Y$, let $f(\pi)$ be the prior probability mass function (PMF) of $\pi$
and $L(\pi \mid y)$ be the likelihood function of $\pi$ given observed data $Y=y$.
Then by Bayes' Rule, the posterior PMF of $\pi$ given data $Y=y$ is
\[f(\pi \mid y) = \frac{f(\pi)L(\pi \mid y)}{f(y)}
\]
where 
\[f(y) = \sum_{\pi} f(\pi)L(\pi \mid y)
\]
$f(y)$ is the overall probability of observing data $Y=y$ across all $\pi$.
It is a constant, thus 
$f(\pi \mid y) \propto f(\pi)L(\pi \mid y)$.
Proportional relationships between the normalized posterior model $f(\pi \mid y)$ for different $\pi$ are preserved 
by the unnormalized model $f(\pi)L(\pi \mid y)$.
Thus, there is no need to compute $f(y)$ for the posterior model \cite{JOD2021}.
Now suppose $\pi\in [0,1]$, the likelihood function $L(\pi \mid y)$ is defined as the conditional PMF
of the Binomial distribution $\rid{Bin}(n,\pi)$:
\[L(\pi \mid y) \equiv f(y \mid \pi) = \binom{n}{y}\pi^y(1-\pi)^{n-y}
\]
and the $\rid{Beta}(\alpha,\beta)$ prior PDF $f(\pi)$ is defined in the standard way:
\[f(\pi)=\frac{\Gamma(\alpha + \beta)}{\Gamma(\alpha)\Gamma(\beta)}\pi^{\alpha -1}(1-\pi)^{\beta -1}
\]
By Bayes' rule, the posterior Beta PDF $f(\pi \mid y)$ is given by
\[
\begin{array}{ll}
f(\pi \mid y) & \propto f(\pi)L(\pi \mid y) \\[0.5em]
& = \frac{\Gamma(\alpha + \beta)}{\Gamma(\alpha)\Gamma(\beta)}\pi^{\alpha -1}(1-\pi)^{\beta -1}\cdot
\binom{n}{y}\pi^y (1-\pi)^{n-y} \\[0.5em]
& = \left[\frac{\Gamma(\alpha + \beta)}{\Gamma(\alpha)\Gamma(\beta)}\binom{n}{y}\right]
\pi^{(\alpha +y) -1}(1-\pi)^{(\beta + n - y) -1} \\[0.75em]
& \propto \pi^{(\alpha +y) -1}(1-\pi)^{(\beta + n - y) -1} \\[0.5em]
& \propto  \frac{\Gamma(\alpha + \beta + n)}{\Gamma(\alpha + y)\Gamma(\beta + n - y)}
\pi^{(\alpha + y) -1}(1-\pi)^{(\beta + n - y) -1} \\[0.75em]
& = \rid{Beta}(\alpha + y, \beta + n - y)
\end{array}
\]
Therefore constructing the posterior Beta PDF involves merely adding to the shape parameters
of the Beta prior PDF $f(\pi)$. 
This also implies the posterior and prior distributions are both
Beta distributions and thus the prior PDF is a conjugate prior for $\rid{Bin}(n,\pi)$.

\end{document}